\documentclass[12pt,titlepage]{article}
\usepackage{amssymb}
\usepackage[dvips]{graphicx}
\begin{document}

\title{Orthogonality criterion for banishing hydrino states from standard
quantum mechanics}
\date{}
\author{Antonio S. de Castro \and \smallskip \\
Universidade de Coimbra\\
Centro de F\'{\i}sica Computacional\\
P-3004-516 Coimbra - Portugal\\
and\\
UNESP - Campus de Guaratinguet\'{a}\\
Departamento de F\'{\i}sica e Qu\'{\i}mica\\
12516-410 Guaratinguet\'{a} SP - Brasil\\
\\
\\
Electronic mail: castro@pesquisador.cnpq.br}
\maketitle

\begin{abstract}
Orthogonality criterion is used to shown in a very simple and general way
that anomalous bound-state solutions for the Coulomb potential (hydrino
states) do not exist as bona fide solutions of the Schr\"{o}dinger,
Klein-Gordon and Dirac equations.
\end{abstract}

An alleged tightly-bound state of hydrogen with strong singularity of the
eigenfunction at the origin (called a hydrino state) has received
considerable attention in the literature \cite{dom}. The order of magnitude
of the atomic size (Bohr radius) as well as the energy of the hydrogen atom
in its ground state just derived from the Heisenberg uncertainty principle,
even in a relativistic framework, should be enough to disqualify hydrino
states. However, in a recent Letter, Dombey \cite{dom} rejects the solution
of the three-dimensional Klein-Gordon equation, previously derived by Naudts
\cite{nau}, as well as the solution of the two-dimensional Dirac equation,
by resorting to a few fair arguments. In addition, Dombey presents a solid
argument founded on the Hermiticity of the Hamiltonian for the Klein-Gordon
case and a suggestion of similar treatment for the three-dimensional Dirac
case. In the wake of Dombey's suggestion, this Letter presents such a
general criterion for banishing hydrino states in the context of the
standard quantum mechanics.

The time-independent Schr\"{o}dinger equation%
\begin{equation}
-\frac{\hbar ^{2}}{2M}\triangledown ^{2}\psi +V\psi =E\psi   \label{1a}
\end{equation}%
and the time-independent Klein-Gordon equation%
\begin{equation}
-\hbar ^{2}c^{2}\triangledown ^{2}\psi +M^{2}c^{4}\psi =\left( E-V\right)
^{2}\psi   \label{1b}
\end{equation}%
with spherically symmetric potentials admit eigenfunctions in the form%
\begin{equation}
\psi =\frac{u_{k}(r)}{r}Y_{lm}\left( \theta ,\phi \right)   \label{rad}
\end{equation}%
where $k$ denotes the principal quantum number plus other \ possible quantum
numbers, $u_{k}$ is a square-integrable function ($\int_{0}^{\infty
}dr\,|u_{k}|^{2}=1$) and $Y_{l}^{m}$ are the orthonormalized spherical
harmonics ($\int d\Omega \,Y_{lm}^{\ast }Y_{\tilde{l}\tilde{m}}=\delta _{l%
\tilde{l}}\delta _{m\tilde{m}}$), with $l=0,1,2,\ldots $ and $%
m=-l,-l+1,\ldots ,l$, in such a way that%
\begin{equation}
H_{\mathtt{eff}}u_{k}=\left( E_{\mathtt{eff}}\right) _{k}u_{k}  \label{auto1}
\end{equation}%
with
\begin{equation}
H_{\mathtt{eff}}=\left\{
\begin{array}{c}
-\frac{\hbar ^{2}}{2M}\frac{d^{2}}{dr^{2}}+V+\frac{\hbar ^{2}l\left(
l+1\right) }{2Mr^{2}} \\
\\
-\frac{\hbar ^{2}}{2M}\frac{d^{2}}{dr^{2}}+\frac{E}{Mc^{2}}\,V-\frac{V^{2}}{%
2Mc^{2}}+\frac{\hbar ^{2}l\left( l+1\right) }{2Mr^{2}}%
\end{array}%
\begin{array}{c}
\quad {\textrm{for the Schr\"{o}dinger equation}} \\
\\
\quad {\textrm{for the Klein-Gordon equation}}%
\end{array}%
\right.   \label{H}
\end{equation}%
and%
\begin{equation}
E_{\mathtt{eff}}=\left\{
\begin{array}{c}
E \\
\\
\frac{E^{2}-M^{2}c^{4}}{2Mc^{2}}%
\end{array}%
\begin{array}{c}
\quad {\textrm{for the Schr\"{o}dinger equation}} \\
\\
\quad {\textrm{for the Klein-Gordon equation}}%
\end{array}%
\right.   \label{E}
\end{equation}

\noindent Meanwhile, the time-independent Dirac equation is
\begin{equation}
H\psi _{k}=E_{k}\psi _{k},\quad H=\vec{\alpha}\cdot \vec{p}+\beta Mc^{2}+V
\label{dirac}
\end{equation}%
where $\alpha _{i}$ and $\beta $ in the standard (or Dirac-Pauli)
representation are given by the 2$\times $2 block matrix form%
\begin{equation}
\alpha _{i}=\left(
\begin{array}{cc}
0 & \sigma _{i} \\
\sigma _{i} & 0%
\end{array}%
\right) ,\quad \beta =\left(
\begin{array}{cc}
1 & 0 \\
0 & -1%
\end{array}%
\right) ,\quad i=1,2,3  \label{4a}
\end{equation}%
and $\sigma _{i}$ are the 2$\times $2 Pauli matrices. Its eigenfunction \
has a spinorial structure given by \cite{gre}%
\begin{equation}
\psi =\frac{1}{r}\left(
\begin{array}{c}
if_{k}(r)\mathcal{Y}_{jm_{j}}^{\kappa }\left( \theta ,\phi \right)  \\
g_{k}(r)\mathcal{Y}_{jm_{j}}^{-\kappa }\left( \theta ,\phi \right)
\end{array}%
\right)   \label{psi}
\end{equation}%
where $\kappa =\mp \left( j+1/2\right) $, with the minus sign for aligned
spin ($j=l+1/2$) and the plus sign for unaligned spin ($j=l-1/2$). Here, $%
\mathcal{Y}_{jm_{j}}^{\kappa }$ (with $j=1/2,3/2,5/2,\ldots $ and $%
m_{j}=-j,-j+1,\ldots ,j$) are the orthonormalized spinor spherical harmonics
resulting from the coupling of two-dimensional spinors to the eigenstates of
orbital angular momentum. The normalization of the Dirac spinor ($%
\int_{0}^{\infty }dr\,\left( |f_{k}|^{2}+|g_{k}|^{2}\right) =1$) requires
that $f_{k}$ and $g_{k}$ are square-integrable functions. Using the identity
$\vec{\sigma}\cdot \vec{\triangledown}=\vec{\sigma}\cdot \hat{r}\frac{%
\partial }{\partial r}-\frac{\vec{\sigma}\cdot \hat{r}\,\vec{\sigma}\cdot
\vec{L}}{\hbar r}$ and \ the properties $\vec{\sigma}\cdot \hat{r}\mathcal{Y}%
_{jm_{j}}^{\kappa }=-\mathcal{Y}_{jm_{j}}^{-\kappa }$ and $\vec{\sigma}\cdot
\vec{L}\mathcal{Y}_{jm_{j}}^{\kappa }=-\left( \kappa +1\right) \mathcal{Y}%
_{jm_{j}}^{\kappa }$, there results that one can write%
\begin{equation}
H_{\mathtt{eff}}\Phi _{k}=E_{k}\Phi _{k}  \label{auto2}
\end{equation}%
where%
\begin{equation}
H_{\mathtt{eff}}=\left(
\begin{array}{cc}
\hbar c\left( \frac{d}{dr}+\frac{\kappa }{r}\right)  & V-Mc^{2} \\
V+Mc^{2} & \hbar c\left( -\frac{d}{dr}+\frac{\kappa }{r}\right)
\end{array}%
\right) ,\quad \Phi _{k}=\left(
\begin{array}{c}
f_{k} \\
g_{k}%
\end{array}%
\right)   \label{Fi}
\end{equation}

It is instructive to examine the radial solutions in the neighbourhood of
the origin for the Coulomb potential $-\hbar c\alpha /r$ ($\alpha $ is the
coupling constant) because $u$, $f$ and $g$ must behave better than $%
r^{-1/2} $ at the origin in order to guarantee their square integrability.
An estimative of the asymptotic behaviour of the radial solutions for small $%
r$ can be obtained by neglecting the terms of order $r^{-n}$ compared with
the terms of order $r^{-\left( n+1\right) }$, where $n=0,1,2$, in the Schr%
\"{o}dinger, Klein-Gordon and Dirac differential equations.

As $r\rightarrow 0$ the terms behaving as $r^{-2}$ ($r^{-1}$) dominate in
the Schr\"{o}dinger case for $l\neq 0$ ($l=0$) in such a way that%
\begin{equation}
r^{2}\frac{d^{2}u}{dr^{2}}-l\left( l+1\right) u=0,\quad {\textrm{for }}l\neq 0
\label{eq2}
\end{equation}%
\begin{equation}
r^{2}\frac{d^{2}u}{dr^{2}}+\frac{2Mc\alpha }{\hbar }\,ru=0,\quad {\textrm{for }%
}l=0  \label{eq1}
\end{equation}%
Substituting the Frobenius series expansion%
\begin{equation}
u=\sum\limits_{n=0}^{\infty }a_{n}r^{\nu +n},\quad a_{0}\neq 0  \label{fro}
\end{equation}%
into (\ref{eq2}) and (\ref{eq1}) yields the quadratic indicial equation
(obtained when $n=0$): $\nu \left( \nu -1\right) =l\left( l+1\right) $,
which has the solutions $\nu _{1}=l+1$ and $\nu _{2}=-l$. It happens that
there is no recurrence formula for \ $l\neq 0$ ($a_{n}=0$ for $n\neq 0$) so
that the general solution for (\ref{eq2}) can be written as $%
u=A_{l}r^{l+1}+B_{l}r^{-l}$. On the other hand, for $l=0$ the power series
expansion fails with $\nu =\nu _{2}$ because only the first coefficient of
the series can be defined. It means that the Frobenius method furnishes only
one solution for $l=0$. A second solution linearly independent of $%
u_{1}=a_{0}r+...$, the solution related to $\nu =\nu _{1}$, can be found by
writing $u_{2}=au_{1}\ln \left( r\right) +\sum\limits_{n=0}^{\infty
}b_{n}r^{\nu _{2}+n}\ $with $a\neq 0$ and $b_{0}\neq 0$. It follows that the
general solution for $l=0$ can be written as $u=A_{0}\left( r-\frac{Mc\alpha
}{\hbar }r^{2}+\cdots \right) +B_{0}\left( 1-\frac{2Mc\alpha }{\hbar }r\ln
\left( r\right) +\cdots \right) $. The above results imply that only
choosing $B_{l}=B_{0}=0$ gives a behaviour at the origin which can lead to
square-integrable solutions. For short, all the normalizable radial
solutions ($u/r$) are regular at the origin and behave as $r^{l}$ as $%
r\rightarrow 0$, for all $l$.

\begin{figure}[th]
\begin{center}
\includegraphics[width=15cm, angle=0]{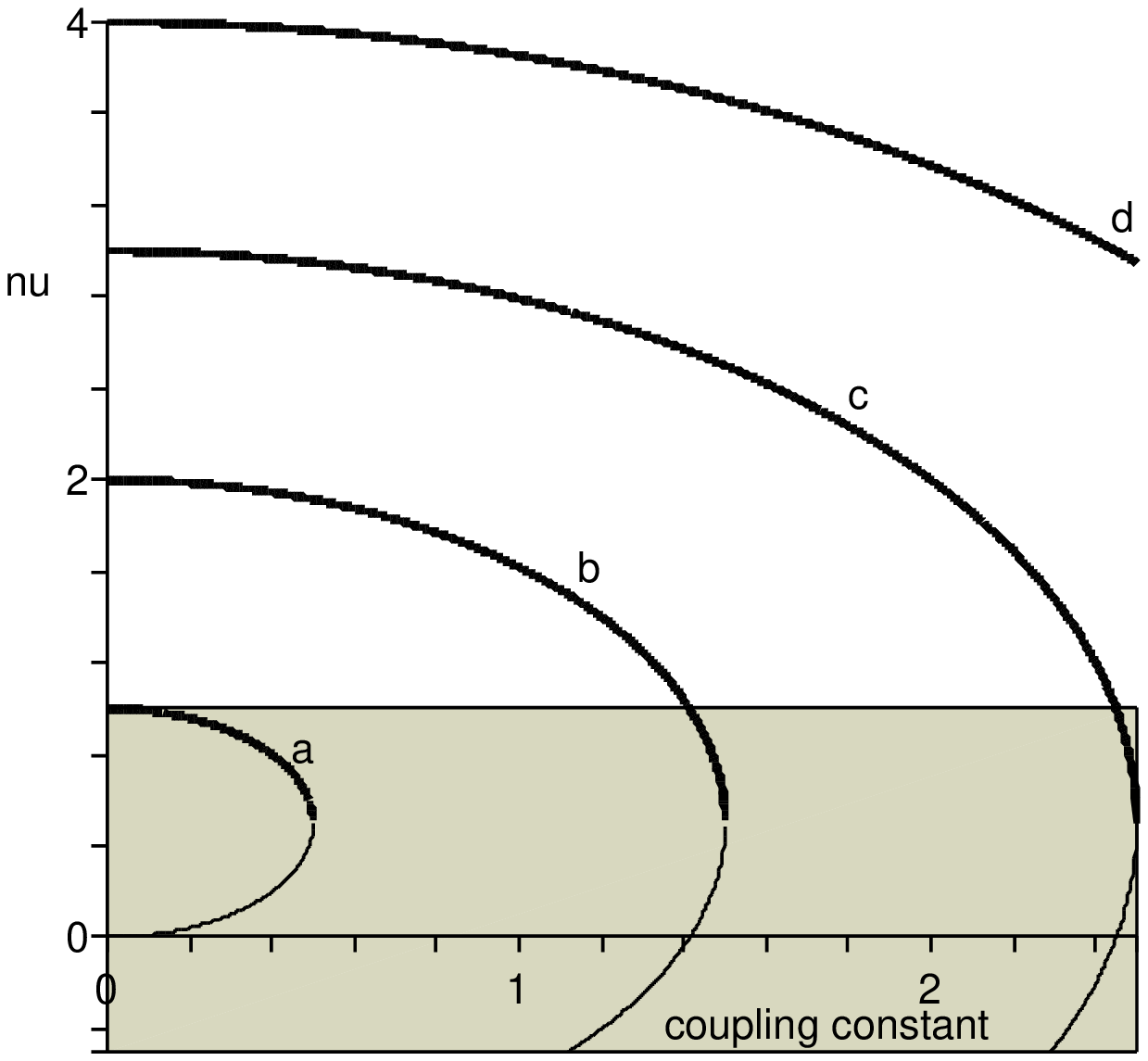}
\end{center}
\par
\vspace*{-0.1cm}
\caption{The roots of the indicial equation, $\protect\nu_{+}$ (thick line)
and $\protect\nu_{-}$ (thin line), as a function of the coupling constant
for the Klein-Gordon case ($a$ for $l=0$, $b$ for $l=1$, $c$ for $l=2$ and $d
$ for $l=3$). The shaded area represents the zone corresponding to radial
solutions diverging at the origin.}
\label{Fig1}
\end{figure}

As for the Klein-Gordon case, the terms behaving as $r^{-2}$, for all values
of $l$, outweighs the other terms  in such a manner that one can write%
\begin{equation}
r^{2}\frac{d^{2}u}{dr^{2}}-\left[ l\left( l+1\right) -\alpha ^{2}\right] u=0
\label{kg}
\end{equation}%
for small $r$. The power series expansion (\ref{fro}) gives the indicial
equation $\nu \left( \nu -1\right) =l\left( l+1\right) -\alpha ^{2}$ which
has the roots%
\begin{equation}
\nu _{\pm }=\frac{1}{2}\pm \sqrt{\left( l+\frac{1}{2}\right) ^{2}-\alpha ^{2}%
},\quad \alpha \leq l+\frac{1}{2}  \label{r1}
\end{equation}%
The argumentation given below is based on (\ref{r1}) and can be better
understood by observing Fig. \ref{Fig1}, where $\nu _{+}$ and $\nu _{-}$, as
a function of $\alpha $, are plotted on the same grid. The constraint $\nu
>-1/2$ which ensures the square integrability of $u$ demands that for $%
\alpha \leq 1/2$ there are just a S-wave \ ($l=0$) solution for $\nu =\nu
_{-}$ $\ $although all the values of $l$ are allowed for $\nu =\nu _{+}$.
Since $\nu <1$ for $\alpha \leq 1/2$, both S-wave solutions, in the sense of
$u/r$, diverge at the origin. As $\alpha $ increases some normalizable
solutions diverging at the origin for $\nu =\nu _{-}$ become possible
whereas some normalizable solutions for $\nu =\nu _{+}$ become divergent. In
any case, as the coupling constant takes critical values ($\alpha
_{c}=l_{c}+1/2$, $\ l_{c}=0,1,2,\ldots $) only the solutions for $l>l_{c}$
are allowed.

\begin{figure}[th]
\begin{center}
\includegraphics[width=15cm, angle=0]{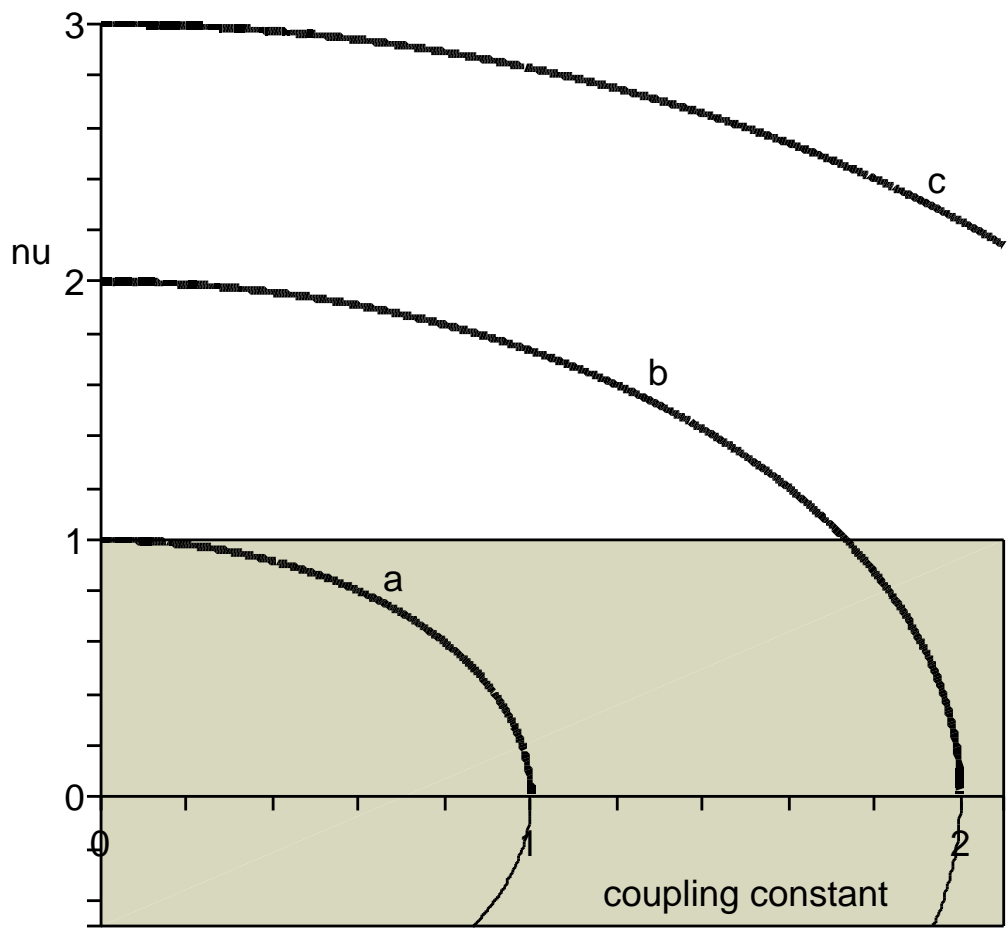}
\end{center}
\par
\vspace*{-0.1cm}
\caption{The solutions of the system of indicial equations, $\protect\nu_{+}$
(thick line) and $\protect\nu_{-}$ (thin line), as a function of the
coupling constant for the Dirac case ($a$ for $|\protect\kappa |=1$, $b$ for
$|\protect\kappa |=2$ and $c$ for $|\protect\kappa |=3$). The shaded area
represents the zone corresponding to radial solutions diverging at the
origin. }
\label{Fig2}
\end{figure}

In analogy with the solutions in the neighbourhood of the origin of the Schr%
\"{o}dinger and Klein-Gordon equations, we look for solutions of the
asymptotic Dirac equation%
\begin{eqnarray}
r\frac{df}{dr}+\kappa f-\alpha g &=&0  \nonumber \\
&&  \label{dirac2} \\
r\frac{dg}{dr}-\kappa g+\alpha f &=&0  \nonumber
\end{eqnarray}%
in \ the form of power series%
\begin{eqnarray}
f &=&\sum\limits_{n=0}^{\infty }a_{n}r^{\nu +n},\quad a_{0}\neq 0  \nonumber
\\
&&  \label{fro2} \\
g &=&\sum\limits_{n=0}^{\infty }b_{n}r^{\nu +n},\quad b_{0}\neq 0  \nonumber
\end{eqnarray}%
Now the coefficient of $r^{\nu }$ ($n=0$) gives the system of indicial
equations%
\begin{eqnarray}
\left( \nu +\kappa \right) a_{0}-\alpha b_{0} &=&0  \nonumber \\
&&  \label{ind2} \\
\alpha a_{0}+\left( \nu -\kappa \right) b_{0} &=&0  \nonumber
\end{eqnarray}%
and the two possible values of $\nu $:%
\begin{equation}
\nu _{\pm }=\pm \sqrt{\kappa ^{2}-\alpha ^{2}},\quad \alpha \leq |\kappa |
\label{sol}
\end{equation}%
Fig. \ref{Fig2} illustrates (\ref{sol}) as a function of the coupling
constant for the lowest values of $|\kappa |$. The requirement of
normalizability ($\nu >-1/2$) implies that for $\alpha \leq \sqrt{3}/2$ only
the positive root is allowed and the radial solutions ($f/r$ and $g/r$) are
found for all the values of $|\kappa |$ ($|\kappa |=1,2,3,\ldots $),
although the radial solutions with $|\kappa |=1$ diverge at the origin. The
negative root allows divergent radial solutions with $|\kappa |=1$ for $%
\sqrt{3}/2<\alpha \leq 1$. Nevertheless, there are no normalizable solutions
with $|\kappa |=1$ for $\alpha >1$. \ Continuing this process, as the
coupling constant increases starting with $\alpha =1$, some normalizable
solutions for $\nu =\nu _{+}$ turn out to be divergent whereas some
normalizable solutions diverging at the origin for $\nu =\nu _{-}$ become
possible. As $\alpha $ takes critical values ($\alpha =|\kappa |_{c}$, $%
|\kappa |_{c}=1,2,3,\ldots $) only the solutions for $|\kappa |>|\kappa
|_{c} $ are allowed. This means that the radial solutions for the smallest
values of $|\kappa |$ become disallowed one after the other as $\alpha $
increases.

In the standard quantum mechanics an observable such as the energy is
represented by an Hermitian operator, whose set of eigenfunctions
constitutes a basis so that every arbitrary wave function can be expanded in
one and only one way in terms of the eigenfunctions. Besides
square-integrability, appropriate boundary conditions must be imposed on the
eigenfunctions of an eigenvalue problem. Square-integrability requires that
the eigenfunction vanishes at the infinity and the boundary condition at the
origin for a singular potential, as the Coulomb potential, comes naturally
into existence by demanding that the Hamiltonian is Hermitian, viz.

\begin{equation}
\int_{0}^{\infty }d\tau \;\psi _{k}^{\ast }\left( H\psi _{k^{^{\prime
}}}\right) =\int_{0}^{\infty }d\tau \;\left( H\psi _{k}\right) ^{\ast }\psi
_{k^{^{\prime }}}  \label{1}
\end{equation}

\noindent where $\psi _{k}$ is an eigenfunction corresponding to an
eigenvalue $E_{k}$. In passing, note that a necessary consequence of Eq. (%
\ref{1}) is that the eigenfunctions corresponding to distinct effective
eigenvalues are orthogonal. Identifying $H_{\mathtt{eff}}$ with $H$ and $E_{%
\mathtt{eff}}$ with $E$ in (\ref{1}), integrating by parts and recalling the
orthonormality of the spherical harmonics, it is easy to show that $u_{k}(r)
$ for the Schr\"{o}dinger and Klein-Gordon cases must satisfy the following
constraint \cite{cas}-\cite{xia}

\begin{equation}
\lim_{r\rightarrow 0}\;\left( u_{k}^{\ast }\frac{du_{k^{^{\prime }}}}{dr}-%
\frac{du_{k}^{\ast }}{dr}u_{k^{^{\prime }}}\right) =0  \label{2}
\end{equation}%
For the Dirac case, identifying $H_{\mathtt{eff}}$ with $H$ in (\ref{1}),
one finds a constraint involving the upper and lower components, namely \cite%
{cas}-\cite{xia}

\begin{equation}
\lim_{r\rightarrow 0}\;\left( f_{k}^{\ast }g_{k^{^{\prime }}}-f_{k^{^{\prime
}}}g_{k}^{\ast }\right) =0  \label{3}
\end{equation}

For the Schr\"{o}dinger case, the square-integrable solution for $u$ ($u{%
\rightarrow r}^{l+1}$ as $r\rightarrow 0$) satisfies (\ref{2}) and its
corresponding radial solution $u/r$ is regular at the origin.

For the Klein-Gordon case, only the solutions for $u$ less singular than $%
r^{1/2}$ can satisfy the orthogonality criterion. One sees, then, that the
square-integrable solution must behave as $r^{\nu _{+}}$, as $r\rightarrow 0$%
, where $\nu _{+}$ is one of the two possibilities of (\ref{r1}),
corresponding to the thick lines in Fig. \ref{Fig1}. Therefore,
square-integrable solutions of the Klein-Gordon equation satisfying the
orthogonality criterion are such that for $\alpha \leq 1/2$ $\ $ all the
values of $l$ are allowed. The $l=0$ radial solution, in the sense of $u/r$,
has a divergence at the origin. As $\alpha $ increases some regular
normalizable solutions become divergent. In any case, as the coupling
constant takes critical values ($\alpha _{c}=l_{c}+1/2$, $\
l_{c}=0,1,2,\ldots $) we are left with the solutions for $l>l_{c}$.

For the Dirac case, only the regular solutions for $f$ and $g$ can satisfy
the orthogonality criterion. This means that the square-integrable solutions
must behave at the origin as $r^{\nu _{+}}$, where $\nu _{+}$ is the
positive square root of (\ref{sol}), corresponding to the thick lines in
Fig. \ref{Fig2}. Here, square-integrable solutions satisfying the
orthogonality criterion are such that, for $\alpha \leq 1$, the radial
solutions ($f/r$ and $g/r$) are found for all the values of $|\kappa |$ ($%
|\kappa |=1,2,3,\ldots $), with the radial solutions with $|\kappa |=1$
diverging at the origin. Nevertheless, there are no normalizable solutions
with $|\kappa |=1$ for $\alpha >1$. \ Continuing this process, as $\alpha $
increases starting with $\alpha =1$, some solutions turn out to be
divergent. As $\alpha $ takes critical values ($\alpha =|\kappa |_{c}$, $%
|\kappa |_{c}=1,2,3,\ldots $) only the solutions for $|\kappa |>|\kappa |_{c}
$ are allowed. This means that the radial solutions for the smallest values
of $|\kappa |$ become disallowed one after the other as $\alpha $ increases.

In summary, a very simple and general criterion has been presented to reject
hydrino states in the context of the standard quantum mechanics. Square
integrability is sufficient enough to exclude singular wave functions in the
Schr\"{o}dinger equation, but not in the Klein-Gordon and Dirac ones. For
the relativistic equations with the Coulomb potential, \ singular wave
functions are allowed. The orthogonality criterion, though, imposes an
additional constraint in such way that the would-be relativistic
square-integrable solutions for hydrino states, related to the thin lines in
Fig. \ref{Fig1} and Fig. \ref{Fig2}, are not acceptable. Therefore, only
radial solutions behaving at the origin as $r^{-1/2+\varepsilon }$, with $%
\varepsilon >0$, for the Klein-Gordon case, and $r^{-1+\varepsilon }$ for
the Dirac case, are physically acceptable solutions.

\newpage

\noindent{\textbf{Acknowledgments} }

The author is indebted to an anonymous referee for very constructive remarks
and to CNPq and FAPESP for partial financial support.

\newpage

\end{document}